# Luminescent nanoparticles in a shrinking spherical cavity – probing the evaporating microdroplets of colloidal suspension – optical lattices and structural transitions

Yaroslav SHOPA [1], Maciej KOLWAS [2], Izabela KAMIŃSKA [2], Gennadiy DERKACHOV [2], Kwasi NYANDEY [2,3], Tomasz JAKUBCZYK [4], Tomasz WOJCIECHOWSKI [2,5], Anastasiya DERKACHOVA [2] and Daniel JAKUBCZYK [2,*]

[1] Faculty of Mathematics and Natural Sciences. College of Sciences, Cardinal Stefan Wyszyński University in Warsaw, Warsaw, 01-815, Poland
[2] Institute of Physics, Polish Academy of Sciences, Warsaw, 02-668, Poland
[3] Laser and Fibre Optics Centre, Department of Physics, School of Physical Sciences, College of Agriculture and Natural Sciences, University of Cape Coast, Cape Coast, Ghana
[4] Institute of Control and Computation Engineering, Warsaw University of Technology, Warsaw, 00-665, Poland
[5] International Research Centre MagTop, Institute of Physics, Polish Academy of Sciences, Warsaw, 02-668, Poland

*Corresponding author: jakub@ifpan.edu.pl

**Abstract**

We investigated the possibility of using charged luminescent nanoparticles as nanoprobes for studying the evolution scenarios of surface and internal structure of slowly evaporating free (light-absorbing) microdroplets of suspension. Three concentrations (1, 10 and 50 mg/ml) of luminescent nanoparticles were used. Single microdroplets were kept in a linear electrodynamic quadrupole trap and the luminescence was excited with a CW IR laser with an irradiance of ~50 W/mm$^2$. Since the microdroplet acted as an optical spherical resonance cavity, the interaction of nanoparticles with light both reflected and modified the internal light field mode structure. Depending on the nanoparticle concentration used, it led, among others, to a very significant increase in modulation depth and narrowing of spherical cavity resonance maxima (morphology dependent resonances – MDRs) observed both in luminescence and scattering, the abrupt changes in the ratio between the luminescence and the scattering and the bi-stability in luminescence signal. The observed phenomena could be attributed to the interaction of optical MDRs with nanoparticle lattice shells forming and changing their structure at the microdroplet surface. In this way, the formation and collapse of such lattices could be detected.

## 1. Introduction

Microdroplets of (colloidal) suspensions are ubiquitous in the environment and technology. When exposed to light, a manifold of phenomena can be observed, in particular when the light is intense and/or the suspension consists of absorbing and/or luminescent nanoparticles (e.g. elastic scattering, Raman scattering, stimulated Raman scattering, stimulated Brillouin scattering, fluorescence, lasing; see e.g. [1] and references therein). Many of the phenomena, in turn, depend on suspended (nano)particles properties (structure) and spatial ordering. In consequence, all these phenomena could be potentially harnessed for remote characterisation of composite microdroplets and diverse studies have been carried out in this field (see e.g. [2–5] for reviews). However, the interaction of light (scattering, luminescence) with composite microdroplets becomes much more complex than for a homogeneous sphere (described with Mie theory, see e.g. [6]). Thus, for example, much effort has been spent on assessing corresponding scattering patterns analytically and numerically – see e.g. [7–10].

For the most part, the structure of resonances corresponding to a homogeneous sphere is gradually supressed with the growing inhomogeneity and finally destroyed (see e.g. [1], compare also [11]). However, the scattering and luminescence spectral properties can also be modified, as the light trapped inside the spherical resonator – the microdroplet – interacts with the suspended nanoparticles (compare [12,13]). Additional resonant features can also arise in scattering and luminescence, when some ordering of the dispersed phase appears, e.g. ordering of nanoparticles into a crystal-like lattice or even just fractal structuring (mostly) at/near the microdroplet surface (see e.g. [14,15]). Such ordering quite naturally arises in an evaporating droplet, in particular when the nanoparticles of the dispersed phase are charged (compare [16–19]). The ordering might also be expected as the result of gradient forces of the (internal) light field (compare laser tweezers [20–22]). A light-absorbing microdroplet can be easily heated (compare [23]) and start to exhibit photophoretic effects as a whole (compare [24,25]) even at relatively low incident light radiant fluxes (like solar radiation [5]). However, at somewhat higher radiant fluxes, as encountered in telecommunication or remote laser probing (compare e.g. [26]), interactions of the internal light field with the dispersed phase in the microdroplet seem quite probable.

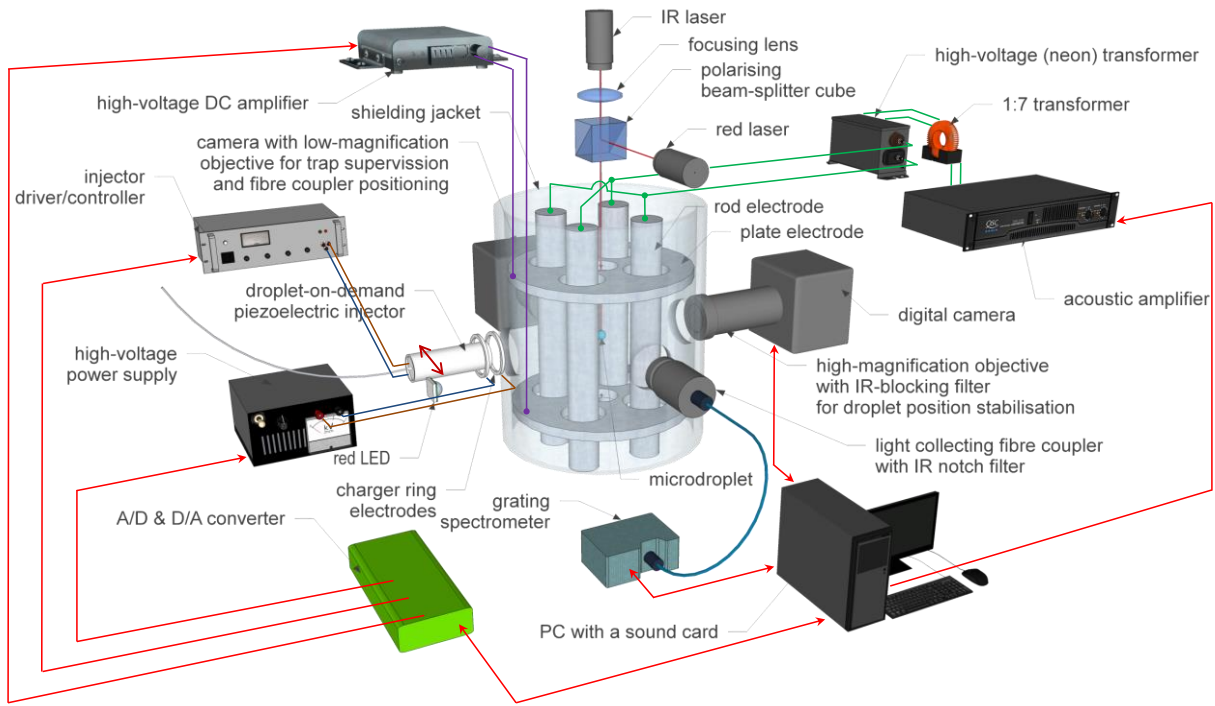

**Fig. 1** Experimental setup schematic visualisation.

The interaction of the internal light field with the dispersed phase could be used as a finer tool for characterization of a composite microdroplet structure and its evolution or, conversely, as suspended nanoparticles manipulation tool (see e.g. [27] and references therein, compare e.g. [28]). In this work, we investigated the applicability of the luminescent nanoprobes for studying the evolution scenarios of evaporating free microdroplets of (light-absorbing) suspension. We used three concentrations (1, 10 and 50 mg/ml) of the nanoparticles and correspondingly analysed (among others): (i) a very significant increase in modulation depth accompanied by narrowing of spherical cavity resonance maxima (morphology dependent resonances – MDRs) observed both in luminescence and scattering, (ii) the abrupt changes in the ratio between the luminescence and the scattering and (iii) the bi-stability in luminescence signal. We attributed the observed phenomena to the interaction of optical MDRs with nanoparticles lattice shells forming and changing structure at the microdroplet surface. In this way the formation and collapse of such lattices could be detected.

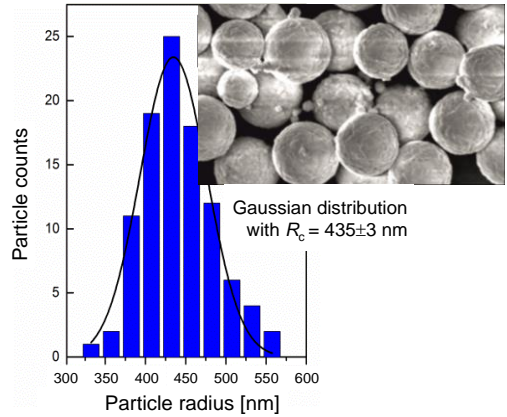

**Fig. 2** Radius distribution histogram of the in-lab produced $Gd_2O_3$:1%$Nd^{3+}$ nanoparticles used in the presented experiments; inset – corresponding Scanning Electron Microscopy image.

## 2. Experimental setup and procedures

The experiments were conducted in a linear electrodynamic quadrupole trap (LEQT) (compare [29–31]) equipped with two plate electrodes perpendicular to the trap axis (Fig. 1). The LEQT axis was kept in the vertical position – an orientation/symmetry favourable for balancing the levitated microdroplet weight and eventually other forces along this axis (see below). Electrically charged droplets were confined horizontally with the quadrupolar AC field of rod electrodes, while vertical confinement was achieved with the DC field of plate electrodes. The field between the plate electrodes in the presented configuration is not homogeneous but exhibits a gradient, which enables the translation of microdroplets vertically by varying the applied voltage ranging from -2.0 to 2.0 kV (depending also on the initial charge of the microdroplet). The levitating microdroplet was backlit with a 1 W red LED and its (vertical) position was observed as a shadow with a digital camera (Smartek, GC651MP, equipped with IR filters) through a high-magnification microscope with a resolution of ~1.0 μm/pixel. It enabled microdroplet radius measurement (shadowgraphy) as well as setting up a stabilization loop and keeping the droplet at the desired vertical location – on the (horizontal) axis of the light-collecting optics. To find this point, an analogue camera with a medium-magnification objective was placed in front of the light-collecting optics. This camera also enabled supervision of the volume of the trap to detect unwanted stray microdroplets /microparticles.

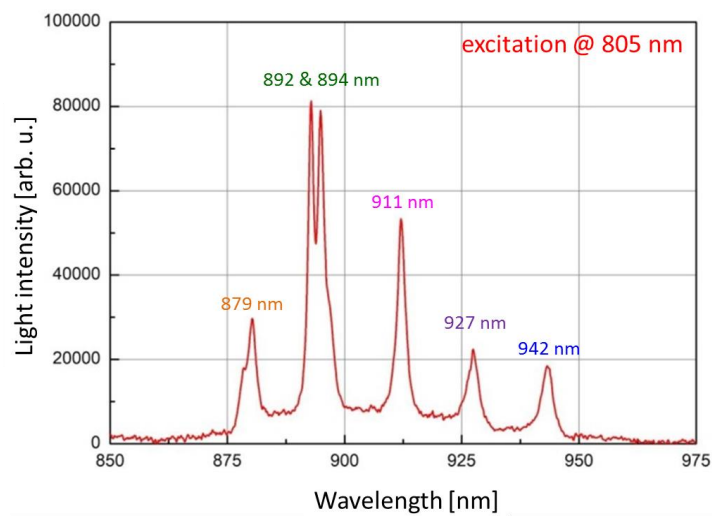

**Fig. 3** Close-lying spectral lines of $Gd_2O_3$:$Nd^{3+}$ nanoparticles luminescence, we measured in a bulk sample with FT spectrometer.

The trap had an aluminium shielding jacket, which screened it from external (static) fields and minimized thermal gradients along the wall. The trap with shielding was kept in an air-tight chamber (not shown in the figure) with windowed ports. All measurements were performed under ambient conditions, while the lab was air-conditioned with temperature set at 22°C.

Microdroplets were delivered to the trap with a droplet-on-demand piezoelectric injector equipped at the nozzle with annular electrodes for droplet charging ([29], compare e.g. [32]). The high voltage used for charging was switched on only for the droplet injection while off directly afterwards, in order not to influence the trapping field during the stabilising loop operation. After the microdroplet injection the injector was moved aside and replaced with the illumination LED.

## 2.1 Sample preparation

In the presented experiments, the microdroplets were formed from a suspension of $Gd_2O_3$: 1% $Nd^{3+}$ luminescent nanoparticles (LNPs) in tetraethylene glycol. The refractive indices were $n_{Gd2O3} \cong 1.98$ and $n_{TetraEG} \cong 1.46$ respectively. The LNPs were produced in-lab with the homogeneous precipitation method (compare [33]). 0.89 g of gadolinium nitrate $Gd(NO_3)_3 \cdot 5H_2O$ and 3.6 mg of neodymium chloride $NdCl_3 \cdot 6H_2O$ were used for the synthesis. 3.6 g of urea $CO(NH_2)_2$ was used as the reducing agent. An aqueous solution (200 ml) of oxidants and a reducing agent was prepared, which was heated to 85°C and shaken in a water bath for 4 h. The obtained nanoparticles were washed in distilled water four times and centrifuged in a laboratory centrifuge. Centrifugation parameters: 6000 rpm, 15°C, 15 minutes. A white powder was obtained which was dried overnight in a drying oven. Next, the nanoparticles were calcined in a laboratory furnace at 900°C for 3 h. We used LNPs with a radius of 435±3 nm (compare Fig. 2). The dispersion medium was chosen for its very low volatility and high polarizability. Importantly, it is also expected that $Gd_2O_3$ nanoparticles in such medium will exhibit substantial electrostatic repulsion (compare e.g. [34]). Slow evaporation of the droplet enabled a long integration time in the light-detection scheme, while high polarizability ensures easy and stable microdroplet charging. The initial concentrations of the prepared suspensions were 1, 10 and 50 mg/ml – below, we refer to it as low, medium/high and high concentration of LNPs. However, due to the density of $Gd_2O_3$ (7.1 g/cm$^3$) being much higher than that of tetraethylene glycol (1.1 g/cm$^3$), sedimentation readily manifests, which may lead to significant variation in the actual initial LNPs concentration in microdroplets. We tried to avoid this by conducting suspension preparation, injector loading and droplet injection (experiment) in quick succession. A single such microdroplet could be trapped in LEQT at 30% RH for up to several hours (compare [35]).

## 2.2 The optical measurement scheme

The levitating microdroplet was illuminated with two perpendicularly polarized CW laser beams of two wavelengths: 805 nm (IR, 1.6 W, p-polarized) and 655 nm (red, 30 mW, s-polarized). Both beams were vertically propagating down the LEQT axis and were used for scattering measurements. The IR beam was also used for luminescence excitation while the red for droplet position stabilization.

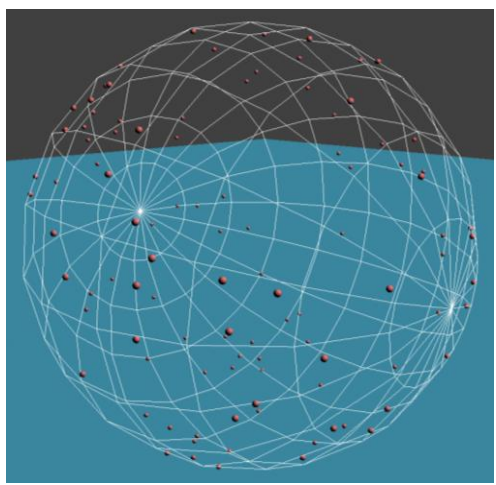

**Fig. 4** Visualization of a distribution of ~100 LNPs in a microdroplet – 1 mg/ml concentration – obtained with the evolution numerical model. The LNP and microdroplet radii were 435 nm and 39 μm respectively. Coulomb repulsion and centrifugal force was taken into consideration, but no interaction with light.

For optical probing of the microdroplet interior, we used $Nd^{3+}$ luminescence at 6 close-lying spectral lines: 879, 892, 894, 911, 927 and 942 nm (Fig. 3). In particular, the 892-894 nm doublet seemed to present an interesting sounding tool. The dedicated experiments on the luminescence of $Gd_2O_3:Nd^{3+}$, conducted with an FT spectrometer (Bomem, DA8) equipped with a PMT (ФЭУ-62) covering the spectral range of 400-1100 nm, showed that the luminescence intensity of $Nd^{3+}$ at the 6 lines mentioned above is comparable to the intensity at 1064 nm – usually the most prominent $Nd^{3+}$ line. For the presented experiments we used a small grating spectrometer with a Si detector (Ocean Optics USB4000, 25 μm slit), which was much easier to integrate with the LEQT.

In order to obtain a measurable luminescence signal, the IR beam was (mildly) focused. In consequence, it could exert significant forces: gradient, photophoretic or both. When the concentration of LNPs was high (~15 mg/ml), the stimulating effect of the IR beam on their mobility in the microdroplet could be observed in the shadowgraphy channel ([Movie 1 in the Supplementary Material](Movie 1 in the Supplementary Material)). Thus the choice of the trap type (LEQT) and geometry/orientation. By following the stabilizing loop DC volt-age, the forces could be balanced and measured, thus providing additional information.

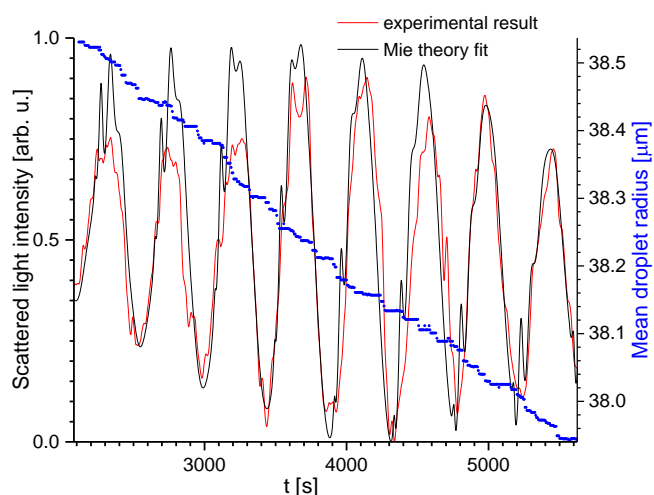

**Fig. 5** The evolution of scattered light intensity at 654 nm for the droplet with low LNPs concentration. Experimental results mildly smoothed (LOWESS, 19/3055 of range window), Mie theory prediction (fit) for a weakly absorbing sphere (see text). Fine resonance structure visible both in experimental and theoretical results, though not exactly fitting. Microdroplet radius (blue dots; 15-percentile filter, 211/958 of range window) obtained with shadowgraphy. Mean evaporation rate – 0.12 nm/s.

The scattered light (red and IR), as well as the luminescence, were collected with a dedicated objective with aspheric achromatic optics. So the radiant flux is integrated over a solid angle with the apex angle of ~15° around the right angle in the scattering plane. We denote this total flux as "light intensity". The objective was equipped with a notch filter (805 nm, OD 6) and coupled to a multimode 600-μm-core fibre feeding the light to a grating spectrometer. Reasonable luminescence signals were obtained for exposure/integration times 0.5-2 s, which set the luminescence and scattering signals' temporal resolution. However, the temporal resolution of the vertical force measurement was limited only by the camera frame rate (~50 fps), which yielded ~20 ms. The evolution of each spectral line intensity was extracted from obtained spectra sequences.

*2.3 Numerical visualization of a microdroplet of suspension*

We have been developing a numerical model for prediction and visualization of dynamics of evaporation-driven aggregation of interacting nanoparticles in a microdroplet [17,36,37]. In this work it was used for visualisation of LNPs distribution at the evolution stage which was of interest to us. (see Figs. 4, 11 and 15). However, we shall briefly describe the model, as it reflects the general notions on the evaporation of a microdroplet of suspension, which we utilise.

The model describes the aggregation phenomenon for the slow-drying regime and considers the particle system to be close to equilibrium. It uses a mixed approach. Various interaction phenomena between the nanoparticles have been accounted for by incorporation of the effective Lenard-Jones (LJ) potential [38], electrostatic interactions and gravitation. Both LJ and electrostatic interaction ranges are limited. There is also a selectable option of imposing a Brownian motion. Movements of each nanoparticle inside the evaporating microdroplet were simulated with the classical Newton's equation of motion. Additionally, to avoid oscillations of colliding nanoparticles, we introduced a dissipative force in the form of liquid viscosity, which conforms to Stokes' law. The radial non-equilibrated forces are introduced to the system by imposing the irreversibly moving surface of the evaporating droplet (gas-liquid interface) dragging up the nanoparticles dispersed in evaporating liquid volume. The interaction of a nanoparticle with the surface (surface tension) was described by a spring constant. The dynamics of the evaporation process can be imposed by analytical formula corresponding to the evaporation rates found in experiments.

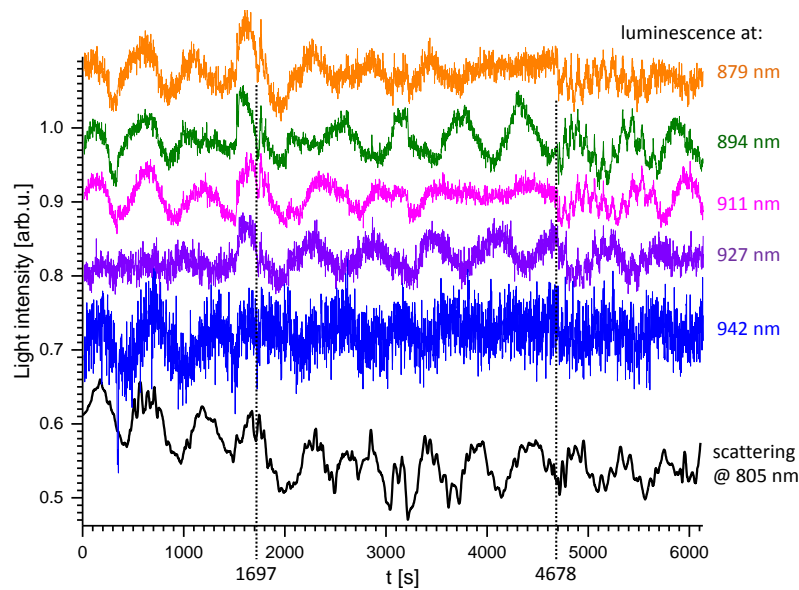

**Fig. 6** The evolution of scattering at the excitation wavelength (mildly smoothed – Savitzky-Golay, 25/3055 of range window) and luminescence intensities corresponding to Fig. 5. The vertical scaling of each graph trace is independent. Increase in modulation depth accompanied by narrowing of resonance maxima visible at 1697 and 4678 s.

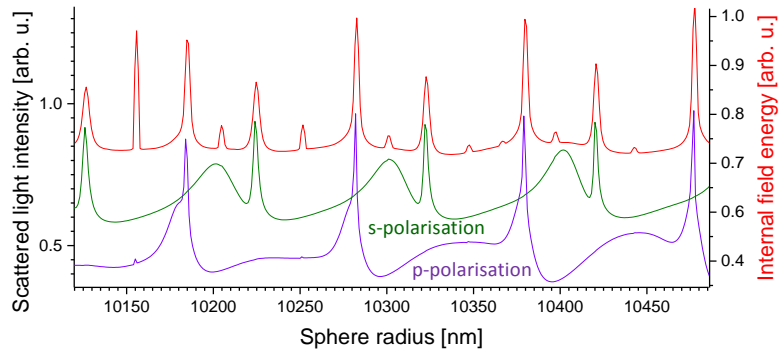

**Fig. 7** Mie theory predictions for a microsphere made of tetraethylene glycol and 805 nm illumination. Integrated far-field scattered light intensity corresponds to experimental observation angle range around the right angle.

The recent version of the code makes use of parallel computing on GPUs with CUDA technology [29]. The aggregation model running the CUDA code allows for a few hundred thousand nanoparticles of several types and is expected to allow for more with the advances in computer hardware. The interactions within each particle type and between types can be freely set. The interaction between a nanoparticle and the surface can be set separately for each nanoparticle type. The centrifugal force associated with the microdroplet rotation was also introduced.

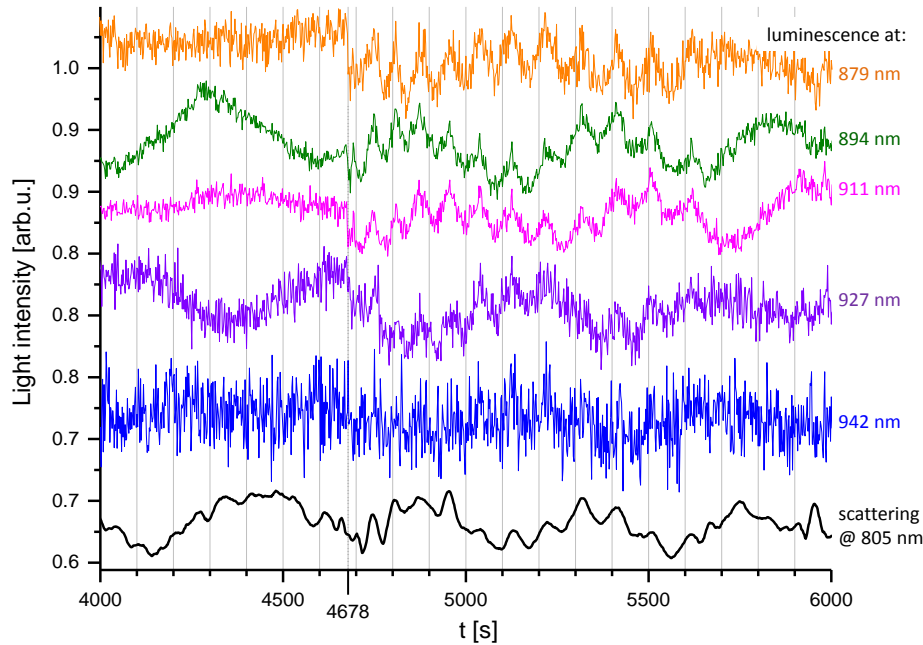

**Fig. 8** The magnification of the final region in Fig. 6. The resonance between the light field mode and the LNPs lattice manifests as the increase of the modulation depth of the fine resonance structure. The synchronisation of this structure visible in luminescence and scattering, the very narrow peaks observable at the centre of the fine maxima, as well as the slight decrease of the luminescence intensity at 4678 s seem to signify the light field assisted ordering of the nanoparticle lattice.

## 3. Experimental results

Depending on the initial LNPs concentration (and microdroplet radius) different behaviours in scattering and inter-droplet luminescence can be observed.

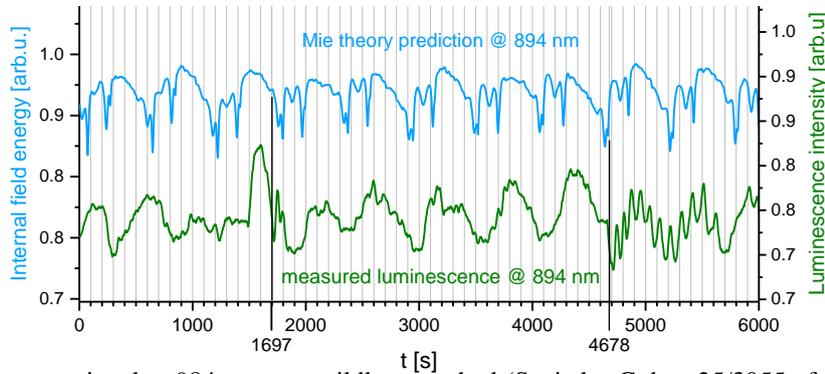

**Fig. 9** Luminescence signal at 984 nm was mildly smoothed (Savitzky-Golay, 25/3055 of range window) to expose the fine resonance structure (compare Figs. 6 and 8). Corresponding total internal field (energy) was generated with Mie theory for $\Im(n)$ found previously in NIR region.

### 3.1 Low LNPs concentration (1 mg/ml)

A very interesting phenomenon could be observed for a relatively low concentration of LNPs. For an initial concentration of 1 mg/ml and a microdroplet of 39 μm radius, there are ~100 LNPs in the droplet (compare Fig. 4). When evenly distributed on the droplet surface, their average separation distance would be ~15 μm. The microdroplet was evaporating very slowly (radius change of ~0.12 nm/s, see Fig. 5) and many optical cavity resonances could be observed. The broad resonances were plainly visible both in scattering and in luminescence all the time. Furthermore, (due to the low light absorp-tion) the resonances modulation depth was generally high. Under closer scrutiny (Figs. 5, 6, 8 and 9), a fine resonance structure could also be detected, again, both in scattering and luminescence. This fine structure is in good agreement with the Mie theory predictions for a homogeneous sphere.

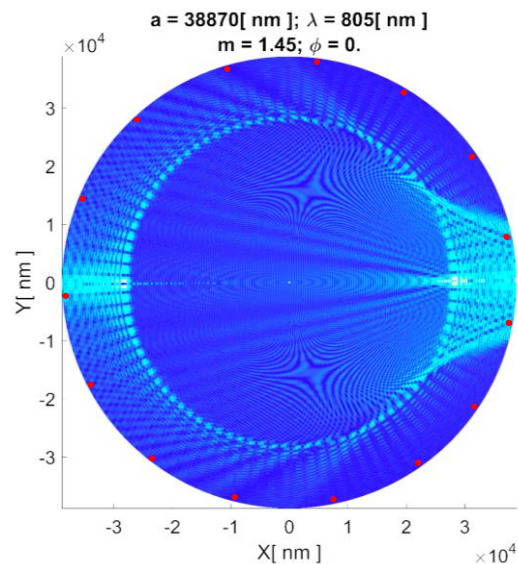

**Fig. 10** Mie theory prediction of internal light field intensity distribution at a MDR for a homogeneous droplet of a radius and refractive index close to the experimental from Fig. 6. Cross section in the scattering plane. Nanoparticles – shown in red – and their regular distribution are in scale, but their position against the field is only indicative.

The broad resonance structure of scattered light intensity at 654 nm could be nicely fitted with Mie theory, while for the fine structure, only some of the peak positions could be reproduced with reasonable accuracy. It is only to be expected since the field modes responsible for the narrow resonances are located close to the microdroplet surface, where LNPs proceed to concentrate and the medium inhomogeneity is highest. The effective imaginary part of the refractive index at 654 nm, estimated with the fitting was $5\times10^{-6}i$.

The luminescence signal was more difficult to interpret but also carried more information. The exact evolution of the luminescence signal is fundamentally very hard to predict, since the location of LNPs is for the significant part random and thus unknown, while the microdroplet usually rotates or tumbles in the trap (compare upper trace in Fig. 9 for homogeneous sphere and e.g. [39]). Interestingly, the fine resonance structure of the luminescence signal consistently corresponded to such structure in the scattering signal. Since the resonances of the internal field usually coincide with those of scattering (see Fig. 7), it can be inferred that the observed correspondence results from the absorption at the higher modes of the 805-nm-wavelength internal field by near-surface LNPs. On the other hand, the broad maxima visible in the luminescence signal – emitted by LNPs in random directions – exhibit the influence of microdroplet eigenmodes at the wavelength of the luminescence (compare Fig. 9). Some synchronization with the excitation field (805 nm) modes seems to persist at the beginning of the droplet evolution.

For the most part of the observed microdroplet evolution, the modulation depth of the fine structure seems in accordance with the $\Im(n)$ found (see Fig. 9; it should be noticed that for 805 nm, $\Im(n)$ is over an order of magnitude higher than for 654 nm). However, at a certain moment (4678 s in the presented case, see Figs. 6, 8 and 9) the luminescence signal slightly but abruptly falls, while the fine structure modulation depth becomes very significant. Furthermore, each constructive fine resonance peak exhibits a very narrow central maximum. The higher modulation depth is also visible in scattering at the excitation wavelength. The fine signal structure is still in synchronicity with the similar structure of the excitation signal (seen via scattering), while the broad resonances are not and correspond to eigenmodes at luminescence lines. The proposed scenario is that the surface LNPs – tending to form a lattice due to their electrostatic repulsion – are further ordered with the forces exerted by the excitation field (gradient or/and photophoretic) at MDR, which direct LNPs to local field minima – hence a slight decrease of overall luminescence intensity. Thus, a very regular LNPs lattice (near the droplet surface)

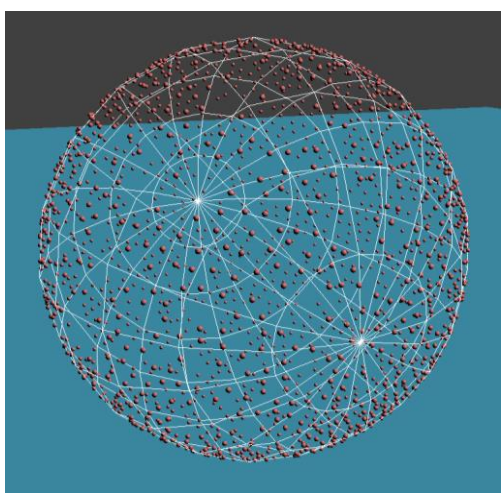

**Fig. 11** A visualization of a distribution of ~1730 LNPs in a microdroplet – 10 mg/ml concentration – obtained with the evolution numerical model. The LNP and microdroplet radii were 435 nm and 46.5 μm respectively. Coulomb repulsion and centrifugal force was taken into consideration, but no interaction with light.

is formed, which is in resonance with the excitation field (modes).

Mie theory prediction of internal light intensity distribution for a homogeneous droplet indicates that suitable light field minima can exist – see Fig. 10 and Movies 2a and 2b in the Supplementary Material (the Matlab code used can be found on GitHub [40]). In the Movies, apart from the internal light intensity also the external near-field intensity distribution up to the distance of 1.5 radius is presented. The sphere radius was decreased in 1-nm and 0.1-nm-steps respectively. The latter was close to the resolution of the experiment discussed in this section. It is worth noticing that for the radius of 38769.9 nm a very narrow resonance (whispering gallery mode) on the background of a wider MDR – not visible at 1-nm-resolution – can be observed. However, we haven't identified such resonances in our experiment with non-homogeneous droplet.

It can be estimated that if each LNP is held with the gradient force of 1 nN (typical of optical tweezers), for the laser power used, the rotation of the microdroplet would be stopped via viscosity forces (from 130 Hz) in a small fraction of a second. The process provides positive feedback (stronger field due to the resonance – finer positioning in the node), which results in the narrowing of the resonance maxima. As long as the MDR lasts, the LNPs stay in the field nodes, but then this ordering is lost. If the MDR is narrower or shallower the manifestation of the phenomenon is shorter, as can be seen around 1697 s (only ~200 instead of ~1000 s).

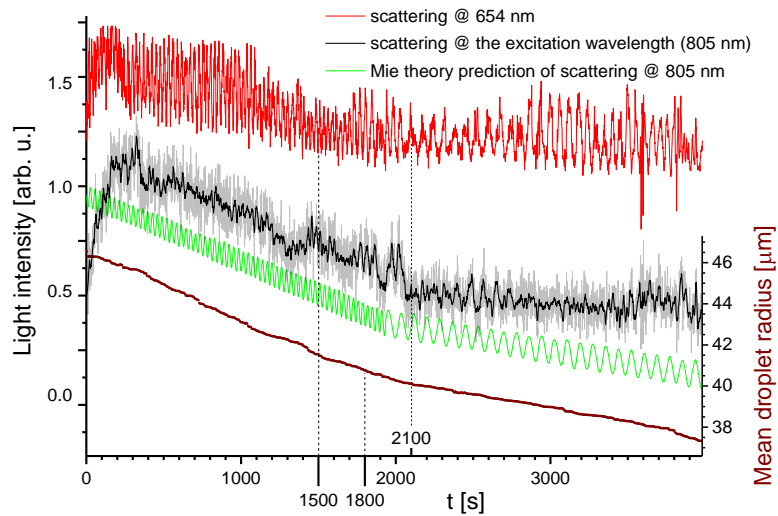

**Fig. 12** The evolution of scattered light intensities and radius for the droplet with medium/high LNPs concentration. Scattering signal at excitation wavelength was mildly smoothed (black line; Savitzky-Golay, 20/4296 of range window). Raw signal is shown in grey. Mie theory-predicted scattering for $\Re(n) = 1.46$, $\Im(n) = 8\times10^{-4}i$ at 805 nm. The vertical scaling of each scattering graph trace is independent. Microdroplet radius (brown dots; 50-percentile filter, 163/699 of range window) obtained with shadowgraphy. Mean evaporation rate changes from 3.4 to 1.4 nm/s.

*3.2 Medium/high LNPs concentration (10 mg/ml)*

For a medium/high concentration of LNPs in the microdroplet, the phenomena plainly visible for the lower concentration are supressed but further can be identified. We shall discuss it on the example of experimental results shown in Figs. 12 – 14. For the initial LNPs concentration of 10 mg/ml and a microdroplet of 46.5 μm radius, there are ~1730 LNPs in the droplet. When evenly distributed on the droplet surface, their average separation distance would be ~4.5 μm. A visualisation of a possible

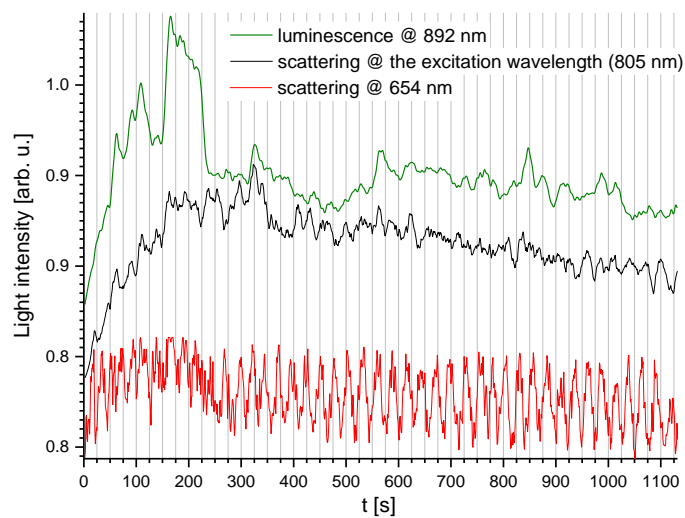

**Fig. 13** The beginning of scattering and luminescence intensities evolution corresponding to the radius range 46.4 – 42.7 μm in Fig. 12. Scattering at the excitation wavelength mildly smoothed (Savitzky-Golay, 20/4296 of range window). The vertical scaling of each graph trace is independent.

LNPs distribution is shown in Fig. 11. In general, the scattering and luminescence signals seem to be more random. It can be attributed to the higher impact of scattering on LNPs, which are more randomly distributed themselves and/or reflect the (random) droplet movements (rotations).

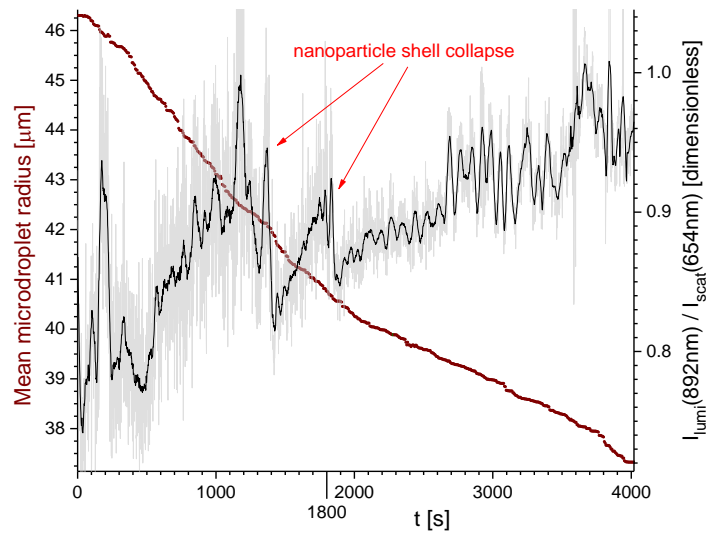

**Fig. 14** Mildly smoothed (black line; Savitzky-Golay, 70/4296 of range window) ratio between the luminescence at 892 nm and the scattering at 654 nm corresponding to Fig. 12. Grey line – the corresponding raw ratio. An increase of modulation depth, characteristic to the formation of a nanoparticle lattice interacting with the resonant cavity (microdroplet) optical modes, can be observed from 600 to 1400 s.

It can be noticed in Figs. 12 and 14 that the evaporation rate changes at ~1800 s. It is plainly visible both as the change of the slope of the mean microdroplet radius evolution measured with shadowgraphy, as well as the decrease of the oscillation frequency of the scattering signal at 654 nm. The process starts at ~1500 s and ends at ~2100 s. In general, such a phenomenon always signifies a change in the microdroplet (surface) composition. In the case of a microdroplet of a relatively dense suspension, it can be expected that an LNPs shell forms, which modifies the evaporation conditions. The shell lattice seems to be governed primarily by Coulomb forces. Further details can be provided by the luminescence signal. Let us notice that at first approximation, the scattering of light on a composite droplet is proportional to its geometrical cross-section (the square of its radius), while the luminescence intensity is also proportional to the number of excited LNPs. Thus, the relative number of excited LNPs can be estimated from the ratio between the luminescence and scattering signals (see Fig. 14). Obviously, the excited LNPs reside close to the droplet surface. We have used scattering at 654 nm, since at 805 nm some nonlinearity may be introduced due to the interaction of LNPs with the light. It can be noticed that, in general, the number of excited LNPs – at the surface – increases until ~1500 s, when an abrupt drop is observed. It can also be observed that before this drop, the signal modulation depth is increasing. A process similar to that discussed in the previous section is expected – a regular LNPs lattice – a dense shell (interacting with the resonant cavity – the microdroplet – optical modes) is formed and then collapses. Some of the LNPs are pushed under, which initially destroys the ordering. Then the process is repeated – a second shell is formed. After that, a significant reduction of the evaporation rate was observed. Further collapses of the shells were not observed in the present-ed experimental run, but the consecutive increase of signal modulation depth is visible, signifying the process of LNPs ordering into a regular structure.

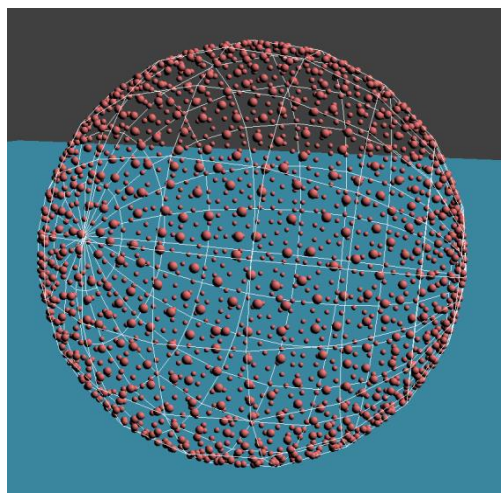

**Fig. 15** Visualization of a distribution of ~1750 LNPs in a microdroplet – 50 mg/ml concentration – obtained with the evolution numerical model. The LNP and microdroplet radii were 435 nm and 27 μm respectively. Coulomb repulsion and centrifugal force was taken into consideration, but no interaction with light.

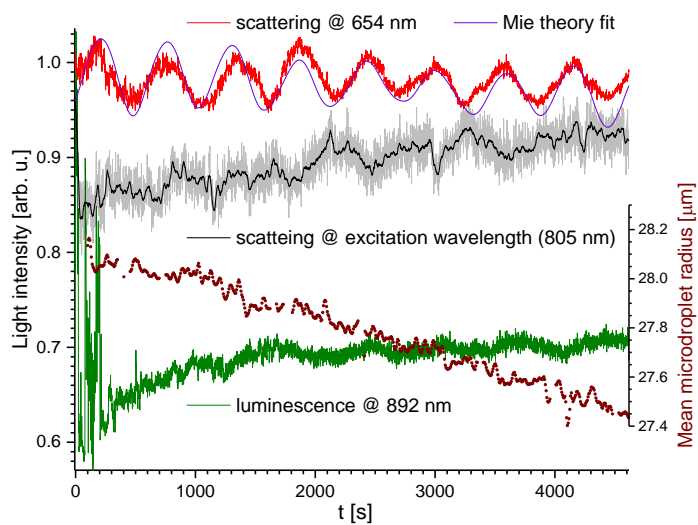

**Fig. 16** The evolution of scattering, luminescence and radius for the droplet with high LNPs concentration. Scattering at 805 nm was mildly smoothed (black line; Savitzky-Golay, 50/5206 of range window). Corresponding raw data is shown in grey. The vertical scaling of each luminescence/scattering graph trace is independent. Mean microdroplet radius (brown dots; LOWESS, 9/1764 of range window) was obtained from shadowgraphy. Mean evaporation rate – 0.14 nm/s.

*3.3 High LNPs concentration (50 mg/ml)*

For a high initial concentration of LNPs, the microdroplet evolution seems to exhibit even more randomness between experimental runs – different evolution paths are possible. A fairly representative evolution sample can be seen in Figs. 16 and 17. For the initial LNPs concentration of 50 mg/ml and a microdroplet of 27 μm radius, there are ~1750 LNPs in the droplet. When evenly distributed on the droplet surface, their average separation distance would be ~2.6 μm. A visualisation of possible LNPs distribution is shown in Fig. 15. As can be seen in Fig. 16, the broad resonance structure of scattered light intensity for 654 nm wavelength (very weakly absorbed) is fairly regular and could be nicely fitted with Mie theory. However, the fine resonance structure is hardly visible (see Fig. 17). Thus it can be inferred that the microdroplet was effectively fairly homogeneous only at larger scales. The effective imaginary part of the refractive index for the given concentration of LNPs at 654 nm, estimated with the fitting, was $8\times10^{-4}i$. The scattering signal at 805 nm is much more irregular and its modulation is hardly visible in general (Fig. 16). It is slightly increasing, similarly as the luminescence, due the increase of LNPs concentration at the surface. All signals oscillate with their corresponding eigenfrequencies, which indicates that the system is not strongly driven at the excitation wavelength for the most part of its evolution. However, at the very beginning of the microdroplet evolution, a certain bi-stability of the luminescence signal can be observed in Fig. 17 – episodes of very high luminescence with very fine and deep modulation. This fine modulation (resonance structure) is in synchronicity with the scattering at the excitation wavelength, very similarly as it was for low con-centration of LNPs (compare Fig. 8). Again, it seems to signify a momentary strong non-linear interaction of LNPs – forming a dense lattice – with the exciting light.

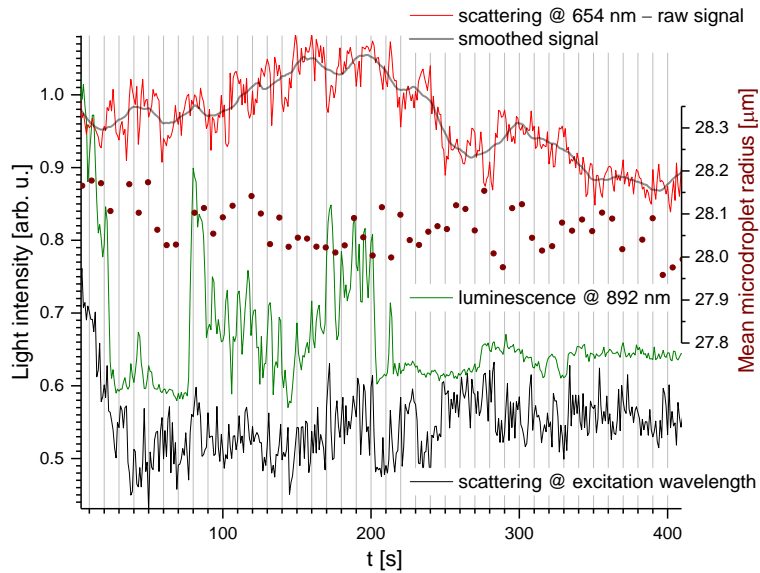

**Fig. 17** Magnification of the beginning of evolutions from Fig. 16 (without Mie theory prediction at 654 nm). Grey line – mildly smoothed (Savitzky-Golay, 45/5206 of range window) scattering at 654 nm shown to visualize the fine resonances structure. Synchronicity of fine resonance structure in scattering at excitation wavelength and in high intensity luminescence regions can be noticed.

## 4. Conclusions

We found that light absorbing (luminescent) nanoparticles in a microdroplet can interact with the internal light field modes of the spherical resonator – the microdroplet. When the irradiance is sufficiently high the nanoparticles not only modify the internal field structure, but can also be directed/ordered by the field. Our previous findings [18] seem to indicate that this can be generalised also to non-absorbing nanoparticles. We found that depending on the nanoparticles distribution, the interaction differs, and its manifestations can serve as a distribution type indicator. Several such phenomena have been identified for different nanoparticles concentration. Most spectacular interaction was observed for the lowest concentration of nanoparticles (1 mg/ml) – optical forces-assisted fine ordering of the lattice leading to a strong resonance between the nanoparticle lattice and the light field mode. It manifested as a very significant increase in modulation depth accompanied by narrowing of spherical cavity resonance (morphology dependent resonance) maxima observed both in luminescence and scattering. For higher concentration of nanoparticles, the formation and collapse of nanoparticles lattice shells at the microdroplet surface was detected – it manifested as the abrupt changes in the ratio be-tween the luminescence and the scattering signals. For the highest concentration of nanoparticles, only the momentary resonances of the regular nanoparticles distributions with spherical cavity modes were observed, manifesting as the bi-stability in luminescence signal – high concentration of random scatterers is known to suppress or disrupt the regular cavity mode structure.

In summary, using luminescent nanoprobes enabled the discovery of the formation (and collapse) of nanoparticle lattice shells on the surface of the microdroplets. However, in some cases, the intense light used for the excitation of the nanoprobes itself induced some ordering of them. Thus, the tool's probing and manipulation capabilities got mixed up to some extent.

## 5. Acknowledgement

This research was funded in whole or in part by National Science Centre, Poland, grant 2021/41/B/ST3/00069. For the purpose of Open Access, the author has applied a CC-BY public copyright license to any Author Accepted Manuscript (AAM) version arising from this submission.